# Neuromorphic Place Cells

Zhaoqi Chen, Ralph Etienne-Cummings

The Johns Hopkins University, Baltimore, Maryland USA

Email: zchen70@jhu.edu

**Abstract**

A neuromorphic SLAM system shows potential for more efficient implementation than its traditional counterpart. We demonstrate a mixed-mode implementation for spatial encoding neurons including theta cells, vector cells, and place cells. Together, they form a biologically plausible network that could reproduce the localization functionality of place cells. The system consists of a theta chip with 128 units and an FPGA encoding 4 networks for vector cells formation that provides the capability for tracking on a 11 by 11 place cell grid. Experimental results validate the robustness of our model when suffering from 18% standard deviation from mathematical models induced by variations of analog circuits. We provide a foundation for implementing dynamic neuromorphic SLAM systems for nonhomogeneous mapping and inspirations for the formation of spatial cells in biology.

## 1   Introduction

Navigation and localization are essential capabilities for animals and humans to survive in complex terrains. As they depart from their homes, they can plan a path through obstacles while advancing towards their targets and remember the path in order to return home. Interestingly, when exploring a new environment, animals can form a concept of the environment that couples spatial location with sensory stimuli to remember locations of, for example, food and danger. This behavior coincides with the scenario of autonomous robots navigating without a pre-charted map. In the robotics community, this problem is called Simultaneous Localization and Mapping [1] (SLAM) and is conceptualized as the computational problem of creating a map of an initially unknown environment and localizing the robot while exploring. However, typical SLAM algorithms require complex computations to handle a large number of sensor inputs and internal memories, resulting in high computational and energy demands. Animals, in contrast, achieve SLAM without explicit mathematical computations and can even navigate in darkness with relatively few sensory inputs. Many researchers [2-6] propose that the neuroscience study of how animal brains navigate and localize would help in developing a more energy- and computation-efficient SLAM algorithm.

The memorizing of environmental information with location implies some form of spatial encoding within the brain. Neuroscience has already observed several classes of neurons whose firing behaviors couple with the animal's spatial location. In 1971, O'Keefe discovered "place cells" in the rat hippocampus; their firing rate is high only when the animal is within a particular spatial location, which defines the associated place field [7]. In 2005, Moser found "grid cells" that activate with spatial periodicity as the animal explores a given space [8]. These results were significant enough to warrant a Nobel Prize in 2014. When encountering an unfamiliar environment, new spatially specific behaviors are generated and persist until further changes of the environment [9-11], demonstrating their mapping capabilities and role in navigation.

Such discoveries have already inspired the development of many biologically inspired SLAM algorithms such as NeuroSLAM and RatSLAM [2-4]. These have shown the advantages of needing fewer sensory inputs and computation power. NeuroSLAM shows promise of performance comparable to that of living organisms, surpassing the current computation-heavy algorithms. Yet problems arise when implementing such models since most of them encode behavior spatially, resulting in complex mathematical computations that must then be implemented in the hardware [12]. This behavior model abstraction fails to take advantage of the elegance and physical properties of underlying neural network circuitry in the hippocampus, resulting in a cumbersome implementation and failing to sufficiently reduce the resource demands to match those typical of a neuromorphic system, whether implemented in software or hardware.

To achieve efficiency like the brain's requires a model of how those spatial encoding neurons—the place cell and grid cells—are generated. Many of the attempted models fall into two main categories: continuous attractor networks (CAN) [13-15] and oscillatory interference (OI) [16-18] models. In CAN models, grid-cell patterns emerge from a network in which each cell has a recurrent connectivity with its neighbors. Each cell has a characteristic preferred direction, meaning that the cell tends to fire more rapidly when the animal is moving in that direction. Each of the cells projects weighted connections, with an off-center circular weight profile, to its neighbors and receives inhibition and excitation from others as well. The local maxima then translate as the animal moves, creating a grid-cell firing pattern.

The OI model Is radically different. It forms grid cells by interfering oscillators, with frequency controlled by the animal's velocity, and encodes the path integration into phase accumulation of the interference results.

Biological observation provides evidence for both models, and both models require a neuron type whose firing activity rises when the animal moves in a certain direction. In 2016, Welday observed theta cells, which have this characteristic [19]. Their firing frequency varies as the cosine of the angle between their preferred direction and the animal's velocity. We will return to this in a later section.

Both models provide challenges for hardware implementation. CAN models are network population models that require large numbers of connections for their recurrent network to function. Furthermore, the weight profile contains excitatory and inhibitory weights that are computed based on distance to neighbors, increasing the computational complexity of a programmable network implemented in hardware. The OI models are much simpler in terms of their modularity since each grid cell receives feedforward excitatory signals from oscillatory cells such as the theta cells. Yet, due to the phase accumulation, OI models require an accurate frequency relationship with velocity and uniformity of its constituent oscillators, which is unrealistic considering the mismatch between either analog VLSI circuitry or biological cells.

The implementation of the fundamental block, i.e., the theta cells, poses difficulties as well. As the attractor model of theta cells suggests that its oscillation can be generated from a group of neurons, organized as ring oscillators with inhibition, to all other neurons in the ring, our team's previous work has shown that it would be difficult to maintain the oscillation's stability while maintaining its sensitivity to velocity inputs [20]. To amend this problem, we designed a theta chip that employs the principle of abstract neomorphism to implement the theta cells' behavior model by exploiting the simplicity of analog computation circuitries. We published an early design for the chip in [21] and then taped out the chip in the TSMC 65nm process. In this paper, we will briefly introduce the final design of the theta chip in section 3, and its performance and proposed application in implementing place cells in section 6.

If a collection of theta cells is available, it is possible to form neuromorphic grid and place cells as the models suggest. But, as mentioned above, both models encounter hardware implementation problems. In the current stage of our research, we choose to implement place cells based on the OI model's framework due to its simpler network structure and single-cell modularity. To overcome its strict requirements on the oscillators, we previously proposed an improved OI model for neuromorphic implementations that compensates the mismatch between theta cell behaviors with an "offset reduction" strategy and simplifies the interference operation to logic operations between oscillators with square-wave oscillation profiles. We describe the principle of the improved model in section 2.2; for a more detailed analysis, please refer to our previous report [22].

Though the model in [22] provides conciseness and feasibility against realistic variations, the spatial cell it forms does not replicate the functionality of a place cell. The model is only capable of indicating one segment of a movement rather than tracking the location of the agent throughout a trail. Thus, in this paper, we will refer to the cells formed by the model of [16] as "vector cells". The original paper proposed a vector accumulation strategy for localization. Here, we realize that concept by proposing a place cell network model that receives input from the vector cells and accumulates the movement in a neuromorphic fashion. We describe the structure of the place cell network in section 5 and how it reacts to implemented vector cell signals in section 6.3.

Thus, in this paper, we demonstrate the results of a hardware implementation of place cells that can track a robot's location. The network for vector cell generation is implemented onto a Spartan-6 FPGA with the theta cell signals from the theta chip. The network structure is computed through our proposed neuromorphic model. The results show that the place cells can be generated from a simple feedforward network with logic node operations from a group of theta cells that are not uniform in behavior. This work proves the feasibility of an efficient neuromorphic SLAM system with place and grid cells generated from a biologically plausible and simple network. In this paper, we briefly illustrate the improved model for place cell generation and describe the design and the characteristics of the theta chip. We demonstrate the theta chip's operation and the implemented model for vector cell generation in the FPGA. After showing empirical results, we discuss the network's path-tracking capability. Lastly, we conclude the findings so far toward identifying further implementation possibilities for theta, vector, and place cells.

## 2 Methods

### 2.1 OI model under ideal conditions

Part of the model that is implemented is an improvement upon the original OI model, tailored for a more robust and simpler implementation in silicon. It can compensate for the inevitable behavioral variations of either biological neurons or neuromorphic implementations while reducing the interference operation to logic operations. The performance and details of the model are published in [22], though we will introduce its basic structure here for completeness. An important characteristic is that to accommodate variations among theta cells readily, the cell formed by the improved model is downgraded into a vector rather than being a location-specific cell in the original OI model. In this section, we first review how to form a place cell in the ideal scenario, then discuss why we can get only a vector cell from either the original OI or the improved model.

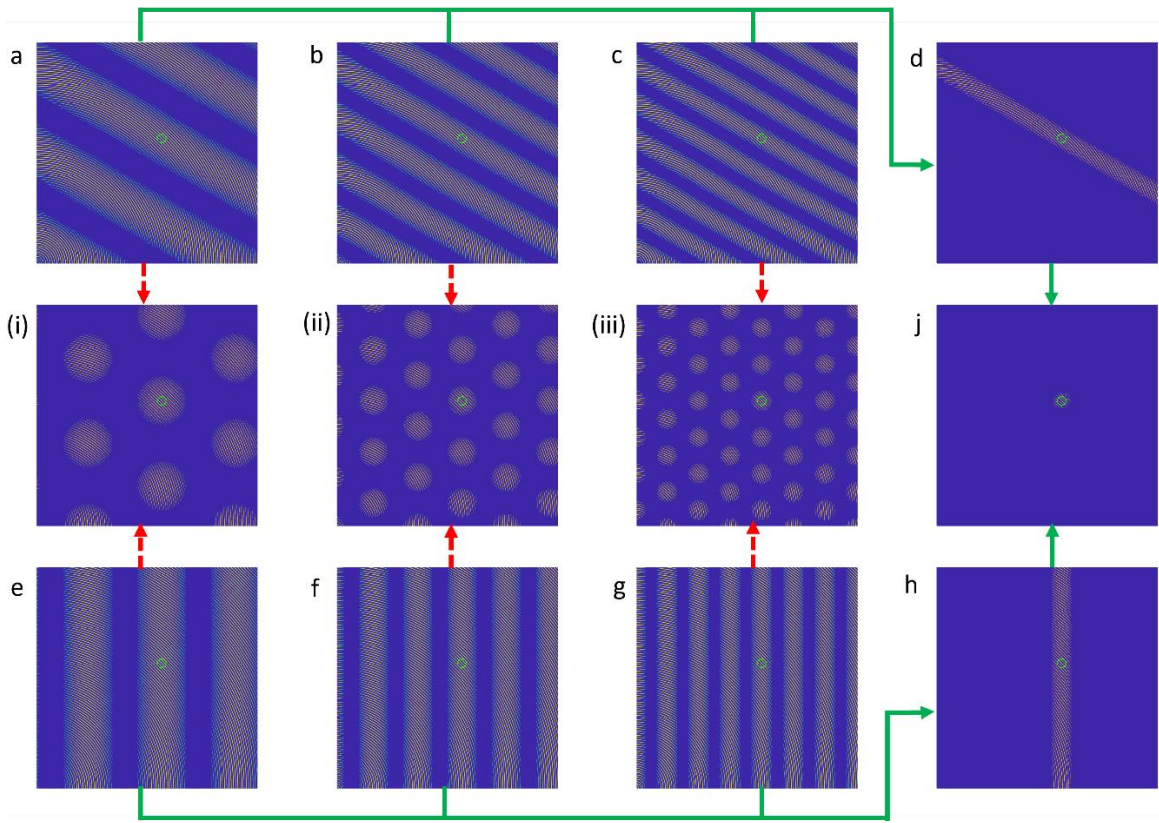

*Figure 1: One method to form a place cell in the ideal scenario: Two groups of theta cells, denoted (a, b & c) and (e, f & h) interfere with a reference theta cell. Both groups have the same set of mutually prime frequency response factors, but with their individual preferred directions 120 degrees apart. A place cell can be formed along the pathway indicated by the green arrow: first, a single stripe is formed, and this is intersected by another at a desired location, marked by the green circle in j. Alternatively, grid cells*

*(i), (ii), & (iii) can be formed by following the red dotted arrows, then interfering the hex grids to form the same place cell in j. This is a reproduction of figure 5 in [22].*

As stated by Burgess et al. [16,17], the fundamental components of an OI model are oscillators that can perform phase accumulation along its preferred direction. The corresponding biological cell type was later discovered by Welday et al., who named them "Theta cells" [19]. Their oscillatory behavior is related to the animals' traversing velocity as described in equation (1).

$$F = F_{\text{idle}} + \beta(\vec{V} \cdot \vec{V_p}) \tag{1}$$

Here, $F$ is the oscillation frequency of a theta cell. It is centered at a constant idling frequency $F_{\text{idle}}$ and modified by a scaled inner product between its preferred direction unit vector $\vec{V_p}$ and the agent's velocity $\vec{V}$. The scaling coefficient $\beta$ is the gain or frequency response factor. The parameters $\beta$ and $F_{\text{idle}}$ are relatively stable for one theta cell but differ between theta cells as Welday et al. discovered [19].

In the OI models, grid cells and place cells can be obtained by interfering with various theta cells [16,17]. First, spatially periodic bands or gratings can be formed by interfering a theta cell with a cell that has an oscillation frequency at $F_{\text{idle}}$, with the results shown in subplots (a,b,c) of figure 1. The process resembles the demodulation of wireless communication to extract the $\beta(\vec{V} \cdot \vec{V_p})$ term that encodes spatial information through phase accumulation over time. The difference in periodicity of subplots (a,b,c) in figure 1 is due to distinct scaling coefficient $\beta$. Grid cells can be formed by interference between gratings of preferred direction that are multiples of 60° apart, as indicated by the red dotted arrows in figure 1. Place cells can be formed in a similar fashion but with gratings with a much larger period, which can be obtained through interference between denser gratings but with mutually prime scaling coefficients, as shown by the green arrows in figure 1.

## 2.2 Improved OI model with theta cell variations

The improved OI model employs a similar interference structure in general but discovered several realistic factors that lead to the improvements and alternation from place cells to vector cells. In a realistic analog implementation or a biological neural circuitry of a theta cell, it is unlikely that all theta cells share the same baseline frequency $F_{\text{idle}}$. Even with a distributed pairing strategy, the chance of finding two theta cells with the exact same $F_{\text{idle}}$ is minimal Here in equation (2), we denote the term $F_{\text{off}_n}$ as the offset frequency for the $n$th theta cell with respect to an average baseline frequency among all theta cells.

$$F_n = F_{\text{idle}_n} + \beta(\vec{V} \cdot \vec{V_p}) = F_{\text{idle}} + F_{\text{off}_n} + \beta(\vec{V} \cdot \vec{V_p}) \tag{2}$$

In the improved model, with schematics on the bottom of figure 2, an offset-reduction strategy is proposed to reduce the impact of this offset term by interfering between theta cells that have similar $F_{\text{off}}$ but opposite preferred directions. We model the logic *AND* interference operation as multiplication here, with the resulting low-frequency component as follows:

$$\cos(2\pi F_a t)\cos(2\pi F_b t) = 1/2[\cos(2\pi(F_a - F_b)t + (\varphi_a - \varphi_b)) + \cos(2\pi(F_a + F_b)t + (\varphi_a + \varphi_b))]$$

$$= 1/2\begin{bmatrix}\cos\left(2\pi[(\beta_a(\vec{V} \cdot \vec{V_p}) + F_{\text{idle}} + F_{\text{off}_a}) - (\beta_b(\vec{V} \cdot -\vec{V_p}) + F_{\text{idle}} + F_{\text{off}_b})]t + (\varphi_a - \varphi_b)\right) \\ + \cos(2\pi(F_a + F_b)t + (\varphi_a + \varphi_b))\end{bmatrix}$$

$$= 1/2\left[\cos\left(2\pi[(\beta_a + \beta_b)(\vec{V} \cdot \vec{V_p}) + (F_{\text{off}_a} - F_{\text{off}_b})]t + (\varphi_a - \varphi_b)\right) + \cos(2\pi(F_a + F_b)t + (\varphi_a + \varphi_b))\right] \tag{3}$$

This operation suppresses the phase error accumulation. Notice that the result has a very similar mathematical form and spatial pattern as the original theta cell, with a new frequency response factor $\beta = (\beta_a + \beta_b)$ and a new offset frequency $F_{\text{off}} = F_{\text{off}_a} - F_{\text{off}_b}$. Furthermore, the summation of frequency response factors $(\beta_a + \beta_b)$ makes the $\vec{V}t$-dependent term in (3) more dominant than the subtracted offset frequency, with the effect shown in subplots (a,b,c) of figure 2. Because of the similarity in behavior, we can view the result as an effective theta cell. This modularity of the offset reduction strategy enables a layered structure as shown in the schematic in figure 2, should the variation be large. This strategy also eliminates the need for idling oscillators and consists purely of theta cells, further reducing system complexity. The variation in $\beta$ among theta cells is less of a problem since it contributes to the uniqueness of the vector cell's firing field, similar to the method of spacing by mutual primes discussed previously. The uniqueness problem is a more predominant issue due to our simplification that converts the sinusoidal oscillation profile of theta cells to square-wave since it basically regularizes the positive part of a sinusoid to one. Please refer to the original publication [22] for a more detailed discussion. In conclusion, the variations in both $\beta$ and $F_{\text{off}}$ becomes advantages when forming a unique firing field for place cell while applying a square-wave oscillation profile, enabling us to use logic *AND* operation for interference to save resources.

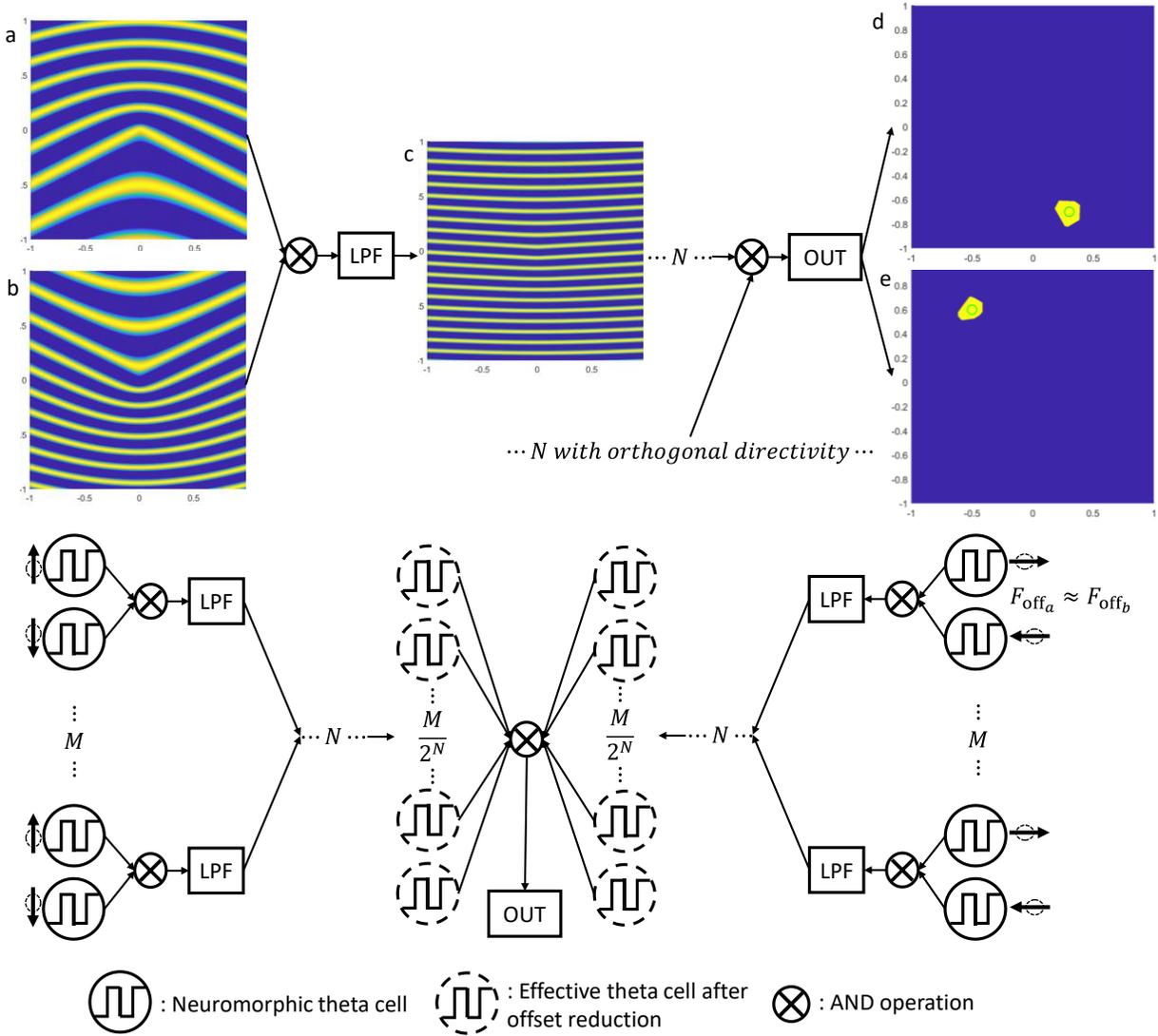

*Figure 2: Subplot a,b,c shows the effect of the offset-reduction method. a&b are the spatial response of two interfered theta cells with $\beta$ = 3.5 and 4.9, $F_{off}$ = 10 and 13.5 relative to the reference frequency, and preferred directions of 0° and 180°. Subplot c is the result of interference between the two theta cells directly; boosting $\beta$ results in a denser grating and less $F_{off}$ to straighten the stripes. Subplots d&e are sample vector cells formed by a group of 48 theta cells with a 25% relative standard deviation on idling frequency and a 20% relative standard deviation of frequency response factor $\beta$. Bottom: Schematic of the offset-reduction method to form vector cells with theta oscillators with variations in $\beta$ and base frequency and non-zero $F_{off}$. Figure from figure 8 of [22].*

In the actual scenario however, we understand that the offset frequency $F_{\text{off}}$ cannot be eliminated all the way to zero, so the phase shift computation must be adjusted as in (4) due to the remaining offset frequency, even after offset reduction. The modulo operation is to account for digitized phase shift in an implemented system, such as a ring oscillator described in the next section that offers 8 digitized phase shifts, or a neural circuit model of theta cell in the form of a ring oscillator [20,26].

$$\varphi_i(R, \theta) = \text{mod}\left(1 - \text{mod}\left(R\left(\cos(\theta - \theta_{\text{p}i})\beta_i + \frac{F_{\text{off}_i}}{|\vec{V}|}\right), 1\right), 0.125\right) \tag{4}$$

Here, $\varphi_i$ is the phase shift of the $i$th effective theta cell after our proposed *AND* interference for the vector cell designated for location $\vec{x}$. $R$ and $\theta$ are the polar coordinates for $\vec{x}$, and $\theta_{\text{p}i}$ for $\vec{V_{\text{p}_i}}$. The spatial firing patterns of two vector cells with arbitrary designated locations are shown in figure 2(d&e). They are formed from the same group of 48 simulated theta cells with a 25% relative standard deviation on idling frequency and a 20% relative standard deviation of frequency response factor.

The reason for the degradation from place cell to vector cell is because of the remaining $F_{\text{off}_i}$ term. They only operate when the theta cells are all in a known state and the agent is traveling with a constant velocity when reaching any point in the spatial maps of figure 2, due to the phase accumulated through $F_{\text{off}}t$. A phase reset procedure is proposed in [22] to address this problem. The procedure is triggered either 1: at a fixed interval when velocity is constant or 2: in the event of changing the movement velocity. The first condition is to clear unforeseen errors, such as thermal noise or crosstalk, accumulated by regularly pulling the system back to a known state. The second condition is due to the dependence of phase accumulated with time due to the remaining offset frequency, generating a path dependency from the origin to any point. For example, the agent can travel from the origin to [1,1] directly, or by detouring via [1,0]. Though both paths end at the same location, the difference in travel times leads to different phase accumulations across the theta cells. Thus, each segment of the path must be recorded as a displacement vector, triggering the phase reset event and creating a new egocentric frame at the point of velocity change. Every time the phase reset event happens, the system can record the current vector cell firing to achieve a path integration or, in other words, a dead reckoning. In section 5 of this paper, we describe our proposed model for place cells to perform localization through the accumulation of vector cell activities.

## 3  Theta chip

### 3.1  Design of the Theta chip

A theta chip has been designed and taped out with TSMC 65nm technology, targeted at producing the behavior described by equation (1). Though a neural attractor network model was previously proposed in [20], it consists of many neurons with both inhibiting and excitatory interconnections, making VLSI implementation and configuration difficult. In order to provide an array of programmable theta cells with reasonable hardware resources, we choose to implement the theta chip with abstract neomorphism by reproducing the theta cells' behavior model. But, to preserve an analogy to the neurological behavior, we employ a mixed-mode design with analog computation circuitry to reduce transistor count as well as a continuous-spectrum response to the input velocity. We also choose digital output for the theta unit oscillations for easier handling by external components. We briefly describe the final design of the chip below; a description of an early design of the included circuits is discussed previously in [21].

The chip consists of 128 theta cell units with an I/O (input-output) arbiter for the control, program, and output of the theta units. Each theta unit has an individually programmable preferred velocity and operates asynchronously. A theta cell includes four major components.

1) SRAMs: Each unit has two 4-bit SRAMs to store the *x* and *y* components of the preferred velocity. The values are signed and in ascending order with value 8 (binary: 1000) representing the zero value. They can be programmed during the chip's start-up phase.

2) A/D Converters: Since the preferred velocity vectors are not to be changed after configuration, capacitance DACs that require refreshment are not preferred here. We adopted a W-2W transistor ladder [27] with a long transistor length to minimize current consumption. Two such DACs are implemented for each theta unit, connecting to the preferred velocity SRAMs.

*3) Analog Computation Module:* The analog computation unit calculates the inner product between the theta unit's preferred velocity and the input movement velocity. We use two Gilbert cells to compute a continuous four-quadrant dot product between the internal preferred velocity and the movement velocity input to the theta unit, both represented in analog voltages. The Gilbert cells require less power and fewer transistors than conventional multipliers. The analog multiplication generates a pair of differential voltages as the product. Two pairs of differential voltages, representing the products of the *x* and *y* components, are converted into two currents using differential pairs, and the sum of the two currents represents the dot product, as shown in figure 3(c).

*4) Oscillator:* The basic structure of the oscillator is a current-starved ring oscillator as shown in figure 3(b), where the control current for the ring oscillator is mirrored from the dot product circuit. An extra NMOS transistor is added between the inverters to allow a manual hold for the oscillation with the control signal Cap_clear, the model needs this phase reset capability for pulling the oscillation to a known phase for correcting accumulated error. To achieve a linear relation between the current and frequency, as well as to reduce the oscillation frequency, extra load capacitance is introduced between inverter stages. Eight of the nine inverters are connected in parallel with a delay capacitor taking advantage of the Miller Effect, thus ultimately reducing the capacitances needed down to 1 pF each. And, since the TSMC technology provides metal-metal capacitors in upper layer metals, the capacitors are placed above the transistors to reduce the area needed for the entire design.

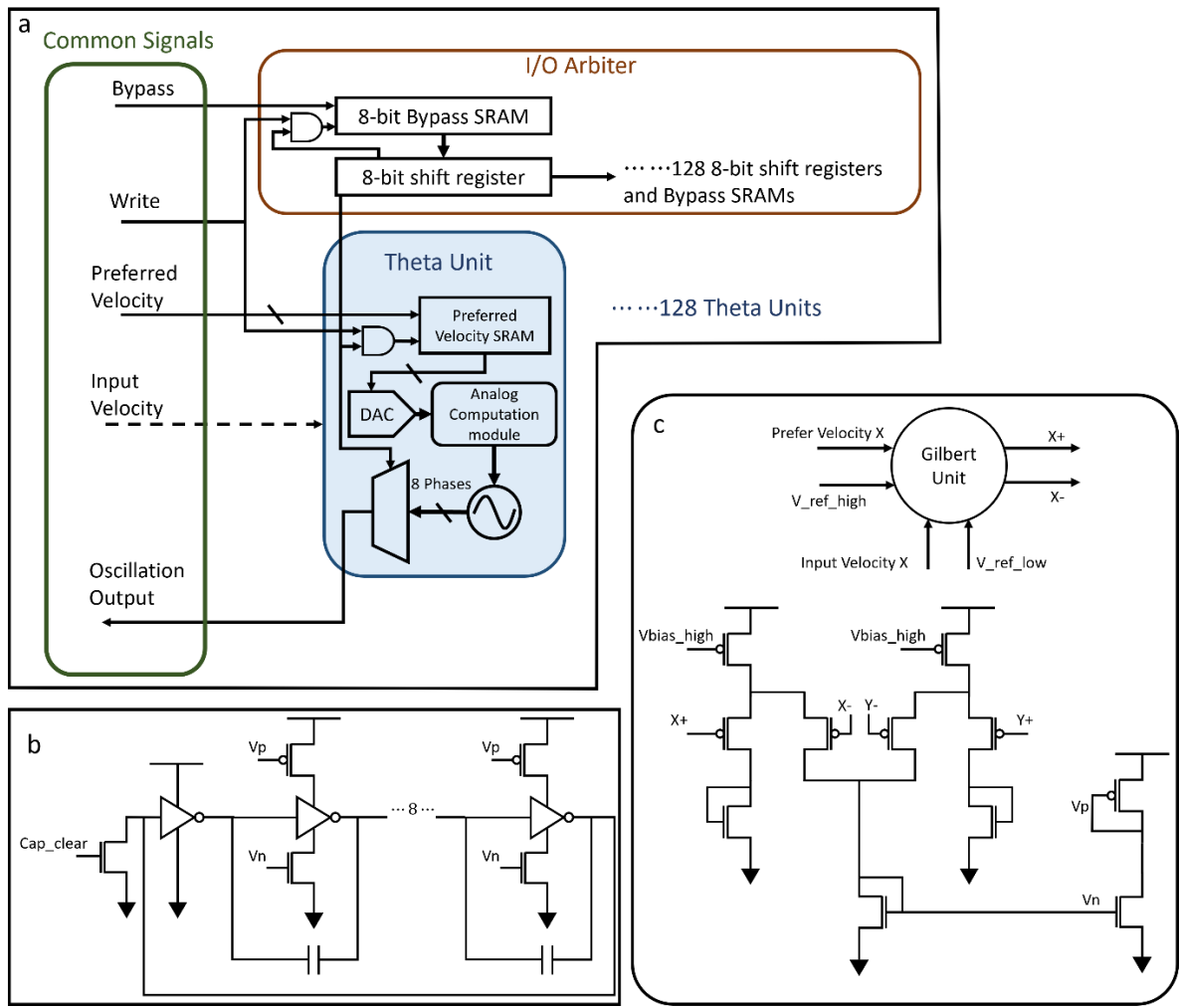

*Figure 3: Theta cell schematics: a. General arrangement of the theta chip. The green "Common Signals" box on the left shows all the broadcasting control and input signals to all theta units, with one shared oscillation output port. A shift-register based I/O arbiter, with bypass capability, is responsible for indexing each theta unit for initial configuration and oscillation output during operation. The theta units store the programmed preferred velocity and perform real-time computation of equation (1) through the*

*analog computation module shown in box c, then regulate the oscillator shown in box b. b: A 9-stage, current-starved ring oscillator with reset capability. c: The analog computation module performs the multiplication using a Gilbert cell and then computes the summation in current mode to achieve vector projection.*

The I/O arbiter follows the concept of TDMA for the I/O of each theta unit, according to the schematic shown in figure 3a. It comprises a chain of 1024-bit shift registers that forms 128 groups of 8-bit registers. Within the group, each bit enables one phase output of a theta unit, and the first bit also provides an `enable` signal for programming the preferred velocity SRAM of this unit. Each register is accompanied by a bypass SRAM that determines whether the shift path goes through the register or bypasses it. The bypass capability allows a shorter scan cycle after the calibration and can only provide signals that are useful for further processing. Upon start-up, the shift register will shift through all the theta units and their 8 phases to program each theta unit's preferred velocity and the Bypass SRAM for each phase output.

### 3.2 Operation of the theta chip

The chip is placed on a custom PCB to interface its digital ports with the FPGA and to meet its analog bias requirement by a DAC. The chip employs serial input and output. Preferred Velocity input is transmitted over an 8-bit bus with the lower 4 bits carrying the *x* component value and the upper 4 bits for *y*. Upon startup, the Clear signal needs to be held high for a few clock cycles to guarantee reset of all SRAMs and shift registers. The first bit of the long shift register will be held high to index the first theta unit's preferred velocity SRAM and the Bypass SRAM for the first phase of the oscillation output of this unit. Then if pulling the active high enabling signal Write, the data input signal Bypass and the Preferred Velocity byte can then be used to program the theta unit's output and preferred velocity. Each clock pulse proceeds the shift register to program whether to output each phase of an oscillation, and every eight clock pulses will move the index to the next theta unit, eventually looping through all 128 units. After the initial setup, the clock then only scans through the phases that are programmed to be output, reducing the length of the scanning cycle or, equivalently, increasing the sampling frequency without increasing the clock frequency. The user can then load the moving velocity of the agent.

| Technology | TSMC 65nm |
| --- | --- |
| Number of Theta units | 128 |
| Maximum output phases | 1024 = 128*8 |
| Die area | 1.5mmx1mm |
| Typical Frequency range | 1206-3066 Hz |
| Typical Mean frequency (Mean/STD) | 2023.771/374.611 Hz |
| Typical of Gain coefficient (Mean/STD) | 20.802/3.688 Hz per unit inner product |
| Digital Power Supply | 3.3V |
| Analog Power Supply | 1V |
| Maximum Scanning Clock | 10MHz |
| Typical Scanning Clock | 6MHz |
| Typical Power Consumption | 6.2mW |
| Output oscillation duty cycle | ~50% |
| Range of input/Preferred velocity [4-bit digital] | Range for better linearity [-4, 4], Max: [-7, 7] |
| DAC linearity | 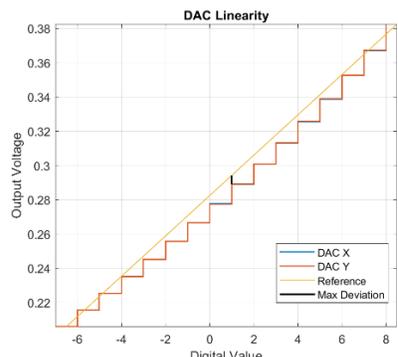 |

*Table 1: Characteristics of the Theta chip.*

# 4 Vector cell implementation

## 4.1 Configuration of the vector cell network

As suggested by the improved OI model, the first layer of interference should be between theta units with similar $F_{\text{idle}}$ but with opposing preferred directions for offset reduction. The total number of layers needed depends on the distribution of $F_{\text{idle}}$. In this demonstration, we implemented only two layers to show the capability of our model with a small number of operations. In figure 8 in the Results section, 82 of the units exhibit a coefficient of determination higher than 0.9. Thus, the structure of a two-layered network will have 40 interference pairs in the first layer and 20 pairs in the second. Among the participating 80 units, units are paired to have the closest possible idling frequencies. The output node will be the direct logic AND operation on the outputs of the 20 nodes of the second layer, with the structure shown in figure 4.

Since the velocity is represented in Cartesian coordinates, the straightforward choice of preferred directions for the theta cells are positive and negative directions along the *x* and *y* axes. Given the programmability of the theta units, we can first form 40 pairs based solely on the closeness of $F_{\text{idle}}$, then assign 20 pairs with opposing velocities along the *x* axis, and the other 20 along the *y* axis. In each pair, for example, one theta unit might be programmed to have a preferred velocity of [4, 0] and the other with [−4, 0]. In the second layer, we can choose to interfere between one from the *x*-aligned group and one from the *y*-aligned group. This operation resembles the formation of grid cells as shown in figure 1(i–iii), showing that one potential functionality of the grid cells is to reduce the influence of offset frequency. An alternative is to repeat the process for pairing based on the similarity of frequencies. Thus, the network consists of 60 interference operations, each of which comprises a logic AND between two signals and a low-pass filtering process as illustrated in figure 4, which is an instantiation of figure 2 but with an inherent formation of grid cells.

## 4.2 Vector cell network implementation

The network for the vector cells is implemented with the Xilinx Spartan-6 FPGA. As shown in figure 4, the implementation comprises two main components: an interference node and a multiplexer. Each node contains simply an AND gate followed by a filter to remove the high-frequency component as suggested by the models. In systems like biological neural networks or similar neuromorphic network systems such as IFAT [28-30], integrators like the neuron's cell membrane or a capacitor for an integrate-and-fire neuron model can perform this low-pass filtering. In our current implementation in the FPGA, we mimic such functionality through a 9-tap Hamming window function followed by a digitized RC circuit. And, to show the compatibility of the system's filters, the filters in the second layer of nodes use a 9-tap moving average filter with a digitized RC circuit with a larger time constant. All the filters are arranged in a pipeline to ensure a continuous data flow and to allow the network output to be synchronized with the clock. The output node, which is simply a 20-input AND gate, produces the output of the place cell which is then buffered in a FIFO queue for streaming to a PC for the next layer, as will be described below.

The multiplexer in figure 4 interfaces between the theta chip and the network described above. For efficiency, the data path structure between network layers is fixed, thus localizing the job of network configuration to the first layer's input. It scans a cycle of the theta chip and directs each oscillation output to its position in an 80-bit wide buffer based on a lookup table, whose values are computed from equation (4) with the frequency information obtained from a calibration cycle. After the multiplexer finishes each scan cycle, it generates a clock signal for the network to load in the 80-bit buffer and advance the interference process.

## 4.3 Interfacing between the theta chip and the network

As described above in section 3.2, the network in the FPGA needs to initialize the theta chip before operation. Based on equation (4), a place cell's designated location is determined by the initial phase shift relation among the theta cells. Combining that outcome with equation (3), every interference result is the difference between the initial phases of the source signals. Thus, in our 2-layered network, the output of the second layer depends on the initial phasing of 4 signals. If we initialize 3 of the signals with the same initial phase, the difference between the initial phase of the fourth theta cell with the other 3 can determine the initial phase of the node of the second layer. So, the number of phases to be output by the theta chip is programmed to be $60 + 8 \times 20 = 220$, where one of every 4 theta units has all 8 phases available for the multiplexer, denoted as the 8 dots around one theta unit in figure 4. After the theta chip is initialized, we gather the traces for the 80 theta units and compute their frequency to compute the phase connection lookup table, given the desired location of a vector cell. In this form, any vector cell (allowing for digitization error) within the range can be generated with the same basis of 220 theta phases by a different lookup table for the multiplexer. Their results for tracking the movement within a certain range are demonstrated in figure 10 in the Results.

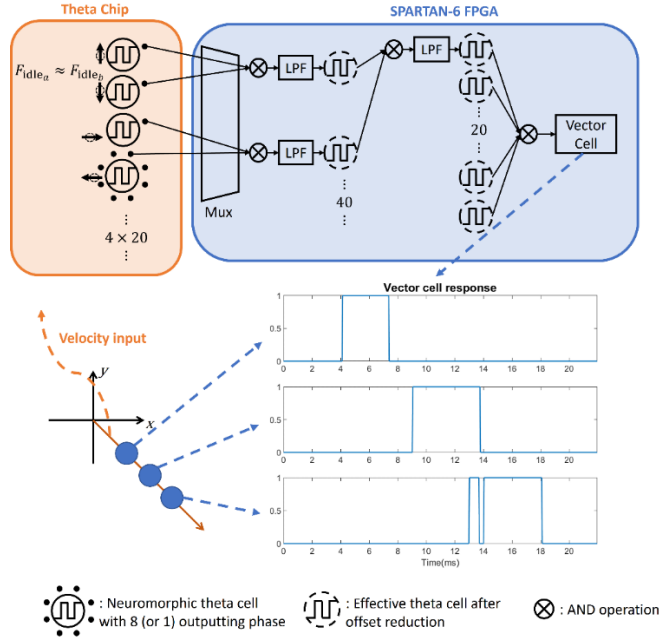

*Figure 4: Implementation schematic of the vector cell network. After a certain velocity is asserted to the theta chip, theta units' oscillations are directed to the input layer of the vector cell network by a multiplexer based on a lookup table. 3 samples of vector cells along the same direction are implemented with their traces recorded in the bottom right figure.*

## 5 Place cell network

Though the vector cells we implemented above can perform basic tracking functionality [22], they do not provide a one-to-one encoding of space as typical place cells are observed to do. And, as stated in [22], a location can be encoded by a certain sequence of firing from such cells. In other words, they function as basis vectors. To realize the spatial encoding capability of place cells, we add another layer of cells that functions as accumulators to represent the vector summation. The structure of this layer is shown in figure 5. Here, a set of self-recurrent place cells are indirectly connected with its neighbors through path cells. The path cells operate like unidirectional switches or AND gates controlled by the vector cells. When a vector cell fires, the path cell that is enabled by an activating place cell is triggered to excite the next place cell while turning off the self-excitatory loop of the current place cell to ensure the transfer of the activity bump. Then, a leakage process is also simulated to discharge the activity of the previous place cell back to baseline level. An alternative way for implementing the transfer of activity bump, especially in the case of actual neural circuits, is to replace the gating of self-excitatory loop by inhibition connections between adjacent place cells. Given the neural firing property of spike rate adaptation, the newly activated place cell would have a spike frequency higher than the previous one thus able to inhibit it back to baseline level thus ensuring the uniqueness of the activity bump after the transfer occurs.

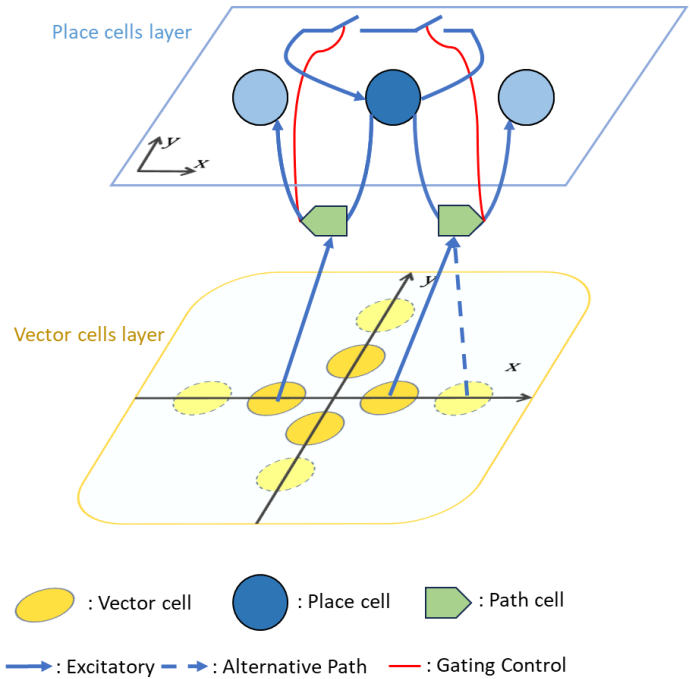

*Figure 5: Place cell network structure for path integration. Each place cell is connected to its neighbors through path cells (y-direction place cells are omitted for clarity) that are gated by the vector cells. When a path cell fires, it transports the activity from the current place cell to its corresponding neighbor and shuts the self-excitatory path off. All sets of path cells can be connected to the same set of vector cells. Here, the path cells are connected to the inner circle of vector cells to create a homogeneous map. A dynamic map can be formed by using vector cells with different magnitudes, as indicated by the faded vector cells and dashed connection.*

This structure loosely resembles the CAN model of place cells in which the localization manifests in the migration of activity bump. The main difference is the lack of direct connectivity between the place cells and the homogeneity among place cells; both drastically reduce the complexity. We believe that the connection between place cells can be used for path planning and navigation but is not necessary for localization, as we have not included it in the model under consideration here, Further description will be postponed until the Discussion. The path cells replace the activity bump transfer functionality that is realized by the asymmetric inhibition profile between place cells and different directional preferences among the place cells in the CAN model. Thus, each place cell is modularized into the structure shown in figure 5. This modularization allows an imp expandable and concise implementation scheme, in contrast to the original CAN model's requiring a substantial number of cells within each patch of the network [13-15]. The path cells outsource the need for directional preference cells to the vector cells which, in turn, can be centralized to save resources that are needed by the reduction of theta cell variations as discussed above and in [22].

Overall, our place cell model greatly simplifies the CAN place cell network by eliminating the complex weight profile between place cells and, thus, the population nature of the attractor networks [13-15]. Moreover, interesting capabilities can be achieved when exploring the connections between path cells and vector cells. For example, if the path cells tap into a different set of vector cells with a different magnitude, the sheet of place cells can be reused to form a space with a different resolution, and similarly for orientation. Furthermore, if we assume learning and plasticity abilities of individual path cells, the map learned by exploration may have a dynamic and inhomogeneous spatial encoding based on the density of the external events and the complexity of the path, opening possibilities for more efficient and dynamic mapping and navigation strategies such as quadtree mapping [31]. As mentioned previously, a phase-reset signal is needed to start the next segment of tracking. In the current setup (figure 5), the reset signal is issued by the firing of any involved vector cells, indicating the end of the current tracking segment.

# 6 Results

## 6.1 Theta chip

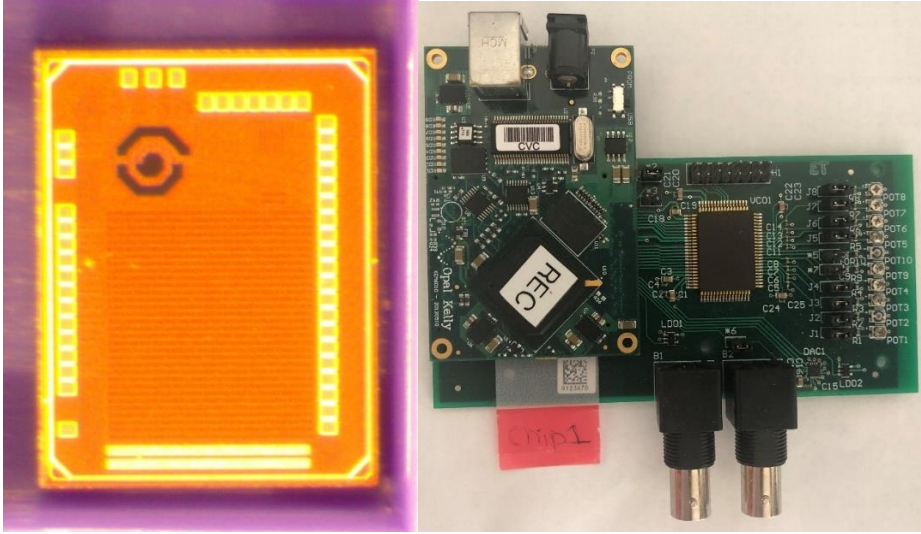

*Figure 6: Left: A photo of the theta chip's die. The left column of pads are digital I/Os and the right-column pads are for analog bias inputs. Right: The top board is a SPARTAN-6 FPGA connected by standard JTAG jumpers to the bottom board where the theta chip sits.*

We first evaluate the theta chip for its accuracy in reproducing the oscillation frequency behavior described in equation (1). To have a more thorough evaluation of the performance, we supplied analog voltages that override the internal DACs for digital inputs. As a few units' responses shown in figure 7, the frequency demonstrates a sigmoid relationship with the inner product value. Thus, during the operation, we chose to limit the magnitude of velocities to approximate the linear relationship. The limitation on the range lead to a shrink in frequency swing as well, allowing us to tune down the current supply for the oscillators without concerning them reach equilibrium on deep negative inner product values. This shifts the idling frequency lower from around 6500Hz to 2000Hz, reducing the power consumption of the chip during actual operation.

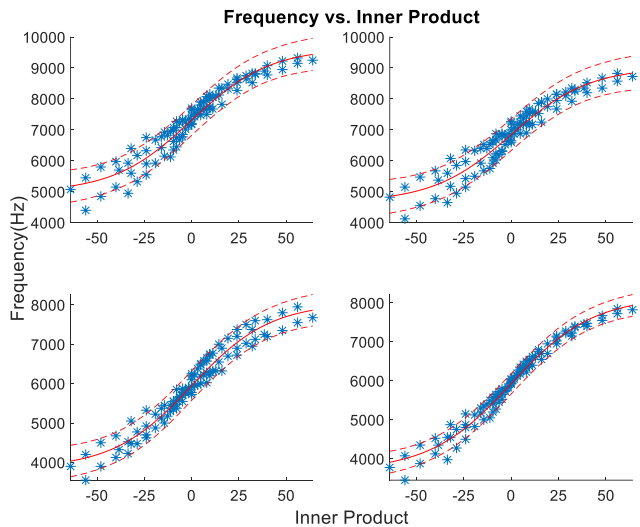

*Figure 7: The response of theta units over a larger range of input velocity values. The frequency depicts a sigmoid relationship with the inner product value.*

The characteristics of the theta chip configured for operation are shown in figure 8. Figure 8a depicts a theta cell's selectivity on the input velocity direction. To assess the performance of the theta chip, we exhaustively program all combinations between the preferred velocity and the input velocity for every theta unit, then record the traces. Frequency analysis then gives a relationship between the inner product of the input and preferred velocities with the oscillation frequencies, with a selection of relevant results with actual configuration values used for vector cell construction are shown in figure 8b. In general, most of the theta units exhibit a positive monotonic relation between the inner product value and the oscillation frequency. Figure 8 shows a scenario in which the input velocity's *x* and *y* components are scanned separately while the preferred velocity is clamped to zero; it shows that some of the variation comes from the W-2W DAC. A slight offset between the zero value of the DAC output and the broadcasting zero-value reference voltage for the mixer resulted in non-zero multiplication results added to the resultant frequency.

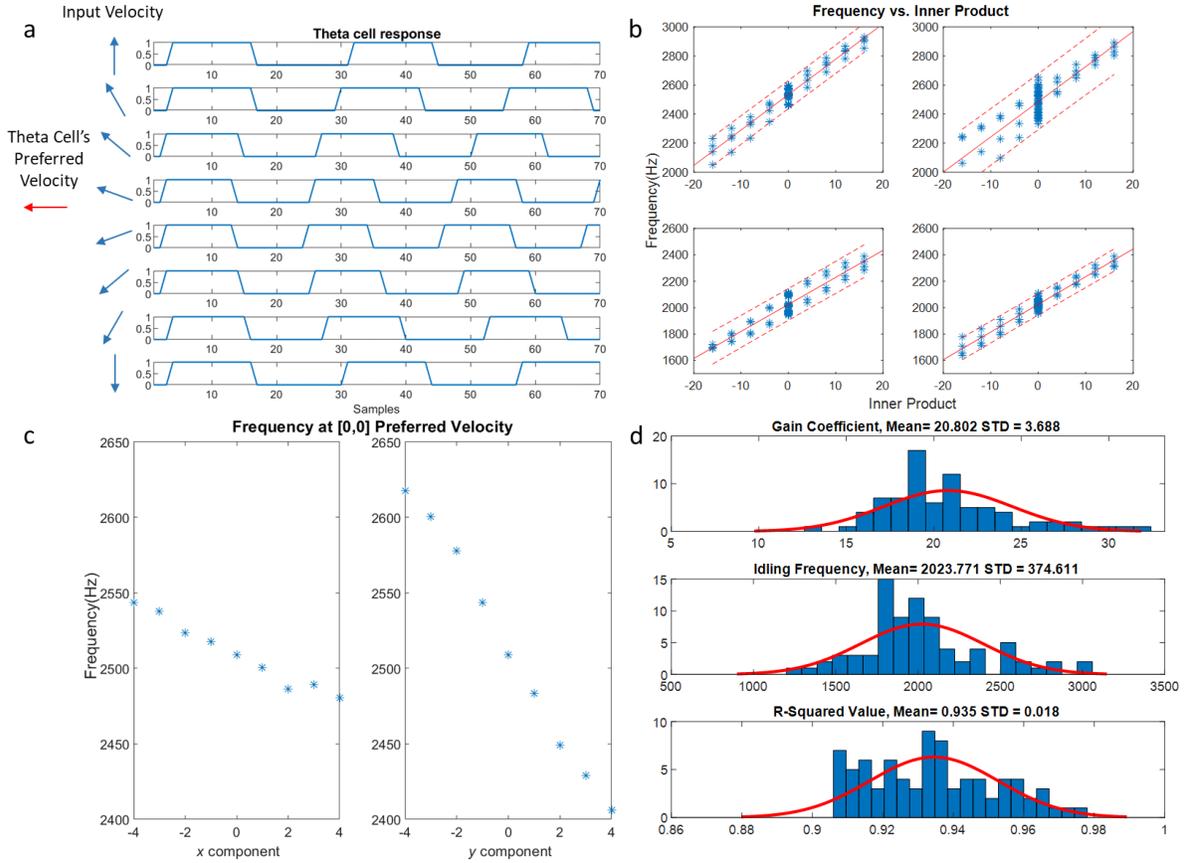

*Figure 8: Theta cell characteristic in operation: a shows traces recorded from a theta cell when the input velocity swings from +90 to −90 degrees with respect to the preferred velocity direction, demonstrating the realization of equation (1). b shows 4 theta units' frequencies with respect to the inner product and the linear fit of equation (1) with 90% confidence bound. The units' preferred velocities are programmed four times with equal magnitudes but opposite directions along the x/y direction, that is further used as inputs to the place cells in the following experiments. They generally depict a linear relationship as expected but have slight offsets for each programmed value. Plot c provides one potential factor for the distribution of frequency for the same inner product value shown in plot b. When the unit's x or y component is programmed to be the zero value (1000 in binary), the frequency still varies linearly with the input velocity's x or y component. This implies that a difference in voltage between the broadcasting zero-value bias and the internal DAC's zero-value output for preferred velocity forms a small value that still participates in the computation. d. depicts the variation in theta cells' parameters after the linear fit to equation (1)*

Given the relationship between preferred velocities, input velocities, and oscillation frequencies of the theta units, we fit the data for each theta chip with equation (1) to evaluate their performances. The results, shown in figure 8, demonstrate variations between the units. Later in the implementation, we choose to only use theta units with a correlation of $R^2 > 0.9$. As shown by figure 8, the values of $\beta$ and $F_{\text{idle}}$ vary considerably among the theta units due to the analog implementation of the computation unit and the oscillator. As described in section 2.2 and in [22], the variation of $\beta$ is in fact beneficial in our model since it substitutes the need to deliberately program parallel preferred velocities to be mutually prime. This variation also avoids the need for a very large frequency domain of operation since the magnitude of programmed preferred velocity can be the same across the units, which helps the oscillators to operate without saturating the current starving capabilities of the ring oscillators. Furthermore, the non-uniformity better resembles the biological behavior of neurons and can, therefore, better test the robustness and plausibility of the proposed improved OI model.

We demonstrate the output flexibility due to the bypass functionality of the chip in figure 9. Although the chip's output is serial, by knowing the number of phases programmed to be output, we could re-parallelize the data stream to reconstruct the oscillation of each outputting phase. Figure 9 demonstrates 8 consecutive phases programmed to be output: the first 4 traces belong to different phases of the same theta unit, and the last 4 traces are from different theta units with different preferred velocities. Based on the measurement, an oscillation frequency range between 1500 Hz and 3000 Hz can produce reasonable linearity with the lowest frequency range. Thus, the Nyquist frequency of the scanning clock becomes $(2 \times 4000 \times N)$Hz, where $N$ is the number of outputting phases. In the experiments, we chose a sampling clock of 6 MHz with 220 outputting phases from the theta chip, resulting in a 27272.7 Hz sampling frequency for each phase.

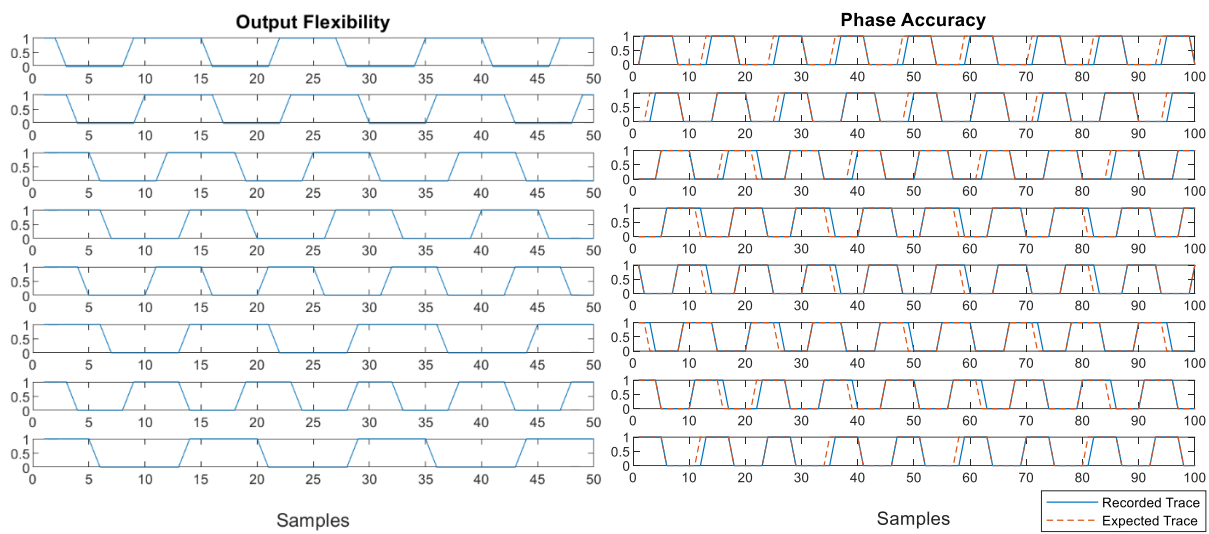

Figure 9: The left figure demonstrates the output flexibility provided by the bypass function of the theta chip through 8 consecutive traces. The first four traces are from different phases of the same unit, and the last four are from different theta units. The right figure compares the eight phases from a single theta unit to a simulated reference trace with $0.25\pi$ phase steps. It shows that, in general, the theta units can provide an accurate initial phase value for the formation of vector cells.

To ensure the formation of vector cells, phase accuracy is crucial. In figure 9, we plot the recorded output of a theta unit's 8 phases and the reconstructed signal with the frequency computed from the Fourier transform, and the respective phase shifts digitized at $0.25\pi$ steps. Most of the units show a good correlation between prediction and output. Adaptive algorithms are also applied when configuring the vector cell networks to ensure the closest fit between the phase tap and the computed phase $\varphi_i$. For cases when the resulted nodes cannot provide constructive interference, eventually, due to cumulative phase alignment error among the participating theta cells, the vector cell network will simply ignore this group.

### 6.2 Vector cell network
An instance of the vector cell network is implemented in the FPGA for validation. We want to test the capability of the vector cells for tracking movements within the domain and the state of aliasing among all the potential vector cells. We generated a grid of 11×11 vector cells as the domain but, due to resource limitations on the FPGA, we could only conduct the experiments for each vector cell within the domain under the same input when building a whole map for the response of all the cells. In other words, the lookup table for each vector cell is reprogrammed for the same input velocity. The start of each session is signified by the release of the `Cap-Clear` signal of the theta chip, then we record the network output. A selection of traces is shown in figure 10, where each square shows whether the value of its corresponding vector cell is high or low. Here, we show snapshots of two trails of the vector cells firing sequentially as the input velocity is held constant. Each vector cell activates only when the agent reaches its place field; this demonstrates its localization functionality. If several of them are deployed simultaneously in combination with the place cell model proposed in section 5, real-time localization can be achieved. Though not shown in the figures, we also observed artifacts from other vector cells coincidentally briefly firing. A similar phenomenon can also be observed in the traces of the relevant vector cells in figure 10, but they are not significant enough to generate a pulse in the subsequent layer which would trigger an activity bump migration, and thus will not affect the tracking accuracy of the place cells.

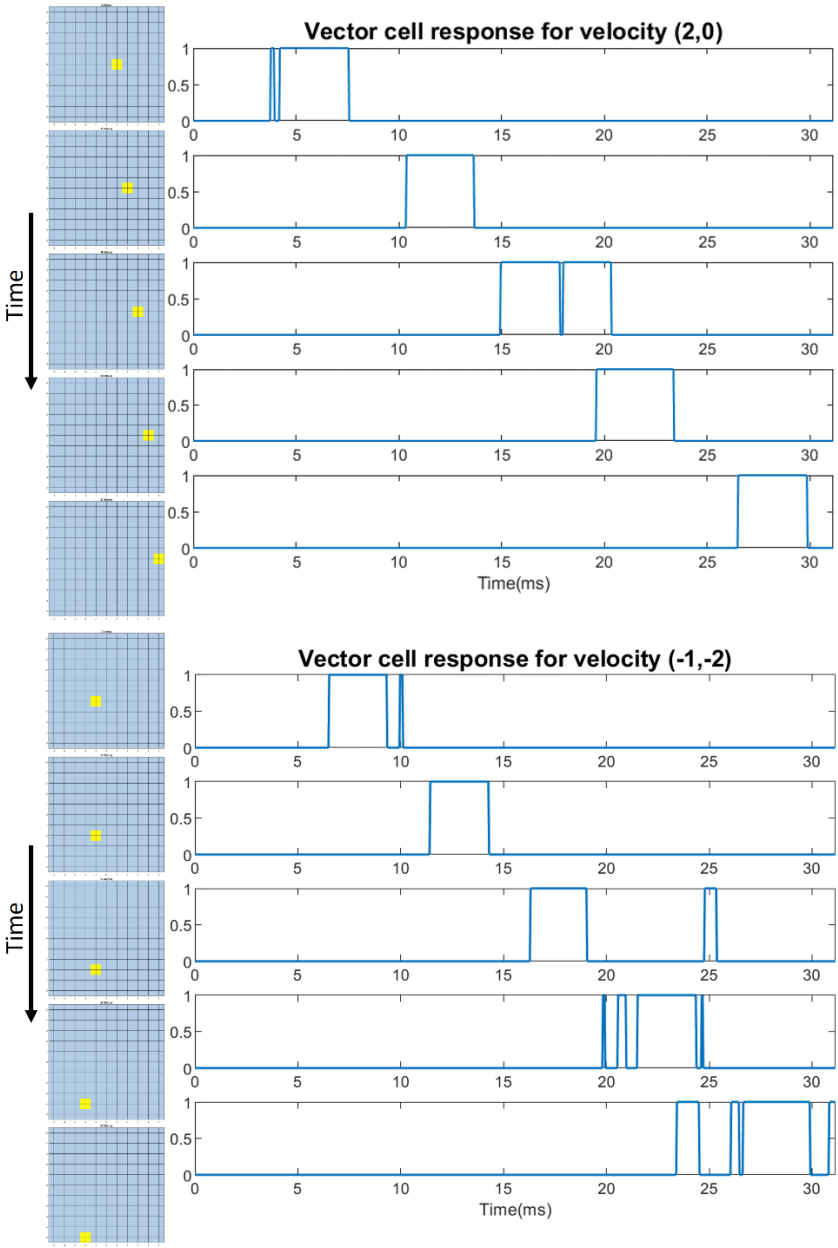

Figure 10: A grid of 11×11 vector cells (with digitization) is generated to test the tracking capability of vector cells. The top two rows of figures show the vector cells firing as time passes, tracking the agent traveling, with their signal traces shown below. We can observe artifacts (short spikes in the traces) as well as random flashes from other vector cells, but they are brief enough not to impact the next layer and the major cause is the limited taps of our lowpass filter.

## 6.3 Place cell network

The location tracking mechanism is powered by the basis formed by the vector cells. Though we discussed the potential advantages of an abundant vector cell basis, the limitation of the FPGA, especially the need for simulated RC filters on board, limits the number of vector cell networks that can be implemented. As shown in figure 4, implementing just one network requires 80 nodes that each involve a 9-tap FIR and a multiplier for the simulated RC filter. But, since the input is digitized, the FIR only requires adders. We expect all these excessive components can be easily replaced by any existing programmable neuromorphic device such as IFAT, or a mixed-mode ASIC chip, after the system is finalized.

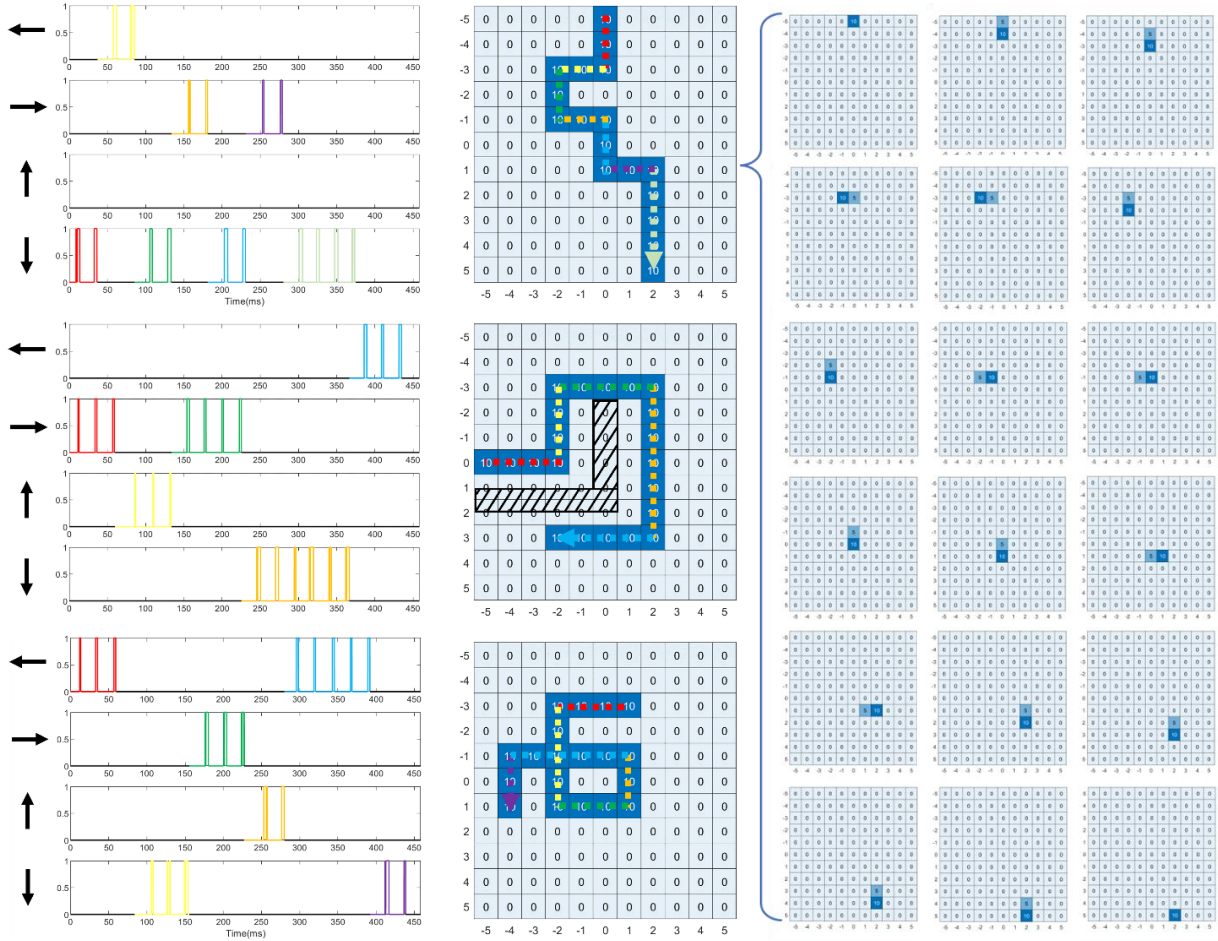

*Figure 11: We simulated a grid of 11×11 place cells to validate its localization function. Three different paths are inputted to the theta chip in the form of velocity sequences. The first of these simulates the agent walking down the arena while looking around. The second simulates the agent circumventing an obstacle to reach its destination at [3, −2]. The third one simulates the agent walking a loop. The traces of the four simultaneously operating vector cells are shown on the left column with colored coded sections corresponding to the movement shown in the middle column. The figures on the middle column mark all the place cells that fired along the trail. The collection of figures on the right shows the place cells in action when receive a pulse from the vector cells, along the first path. The activity bump marked by a value of 10 moves to the next place cell while leaving a trailing tail of 5 because of the global leakage of 5.*

We chose to implement the conventional Cartesian basis on the cardinal directions (N, S, W, E), so 4 such networks need to be instantiated. However, on close observation, due to our 2-layered structure requiring only 1 out of 4 theta cells to have a programmable phase, half of the first layer's nodes are constant and thus can be shared with other vector networks to save resources. In general, the number of sharable nodes is:

$$S = \sum_{n=1}^{N-1} \frac{2^n - 1}{2^N} M = M\left(1 - \frac{N+1}{2^N}\right) \tag{9}$$

Here, $S$ is the number of sharable nodes, $N$ is the number of layers and $M$ is the number of theta cells involved. Thus, the total number of nodes for $K$ vector cell networks is

$$Q = M\left(1 - \frac{N+1}{2^N}\right) + KM\frac{N}{2^N} = M\left(1 + \frac{(K-1)N - 1}{2^N}\right) \tag{10}$$

The number of nodes or resources needed for multiple networks becomes less significant as the number of layers increases. In the current setup, we could use $20 + 4 \times 40 = 180$ nodes instead of $4 \times 60 = 240$ for the implemented basis, albeit needing more DSP capacity than is provided by the SPARTAN-6 FPGA used here.

To conduct a tracking experiment, commands for controlling input velocity and resetting to the theta chip are issued from the host PC to the FPGA, then the PC records the output from the vector cells and feeds the vector cell traces into the place cell network. The `Cap-clear` signal is generated at the beginning of the trail and when any vector cell fires, or a change in input velocity. Currently, for ease of display, the place cell network is simulated in MATLAB. It is trivial to implement the place cell network in its current state on a FPGA or a general-purpose neuromorphic chip given the modular design and simple logic. But we believe it can be further developed to perform mapping and navigation functionalities with the interconnection between place cells; this prospect will be explored in the Discussion.

Three of the experiments in figure 11 demonstrate tracking capability with the implemented vector cell networks receiving inputs from the theta chip. The first path simulates the agent traveling along while randomly looking around. The second simulates the agent circumventing an obstacle to reach a certain destination, and the third tests the place cell network's ability to handle a trajectory with a loop. We use 5 as the step size for each place cell's activity level to allow clearer differentiation. A value of 10 means that a place cell is currently firing; 5 means that it was firing then decayed by a universal leak parameter of 5. We use this scheme to create a trailing tail for a better demonstration of the travel direction, shown in figure 11. We can see that the place cell network can successfully accumulate the displacement vector signals to become a tracking map in an absolute spatial location given any path.

## 7  Discussion

We have demonstrated a neuromorphic structure for the formation of place cells with a simple digitized interference network that shows localization and path tracking capabilities in a place cell network. Our theta chip shows variations just like those that Welday et al. observed in biological theta cells. We validated the robustness of the OI model proposed in [222] with our theta chip and FPGA implementation. We also designed a simple place cell network to convert the vector information of the OI model in [2] into a functional localization system.

The proven effectiveness of the improved OI model opens possibilities for neuromorphic implementations of a neuromorphic SLAM system. The model's tolerance towards theta cells allows for an analog implementation for lower layers. Our analog structure for the behavior model of theta cells serves as an example. Other theta cell implementations with more biologically plausible structures can also be developed through our structure without strict restrictions on accuracy or linearity. Furthermore, variability in the frequency response factor $\beta$ can naturally reflects the uniqueness of biological vector cells.

Our theta chip functions as a valid input layer despite the fact it operates at a much higher frequency range (~2500 Hz) than is observed in biology (~8 Hz). This issue shortens the phase-reset period, thus reducing the operation domain for the vector cells. We will consider reducing the operating frequency of the theta chip in our next iteration.

The improved OI model provides a potential explanation of the functionality of grid cells for localization. The nodes in the second layer of our network structure can form grid cell behavior as shown in figure 1. Thus, grid cells not only participate in the localization, but also serve the purposes of variational reduction. In the second layer, there are two possibilities of pairing. One, shown in figure 1(i–iii), is the pair formed by lower-layer nodes with different orientations that generate grid cell patterns such as the red arrow in the figure. The other possibility is that the pair is formed by nodes with the same orientation, resulting in a sparse stripe pattern as shown by the green arrow in figure 1(d, g). Interestingly, both types of cells have been observed in biology [25]. Our model can serve as a guide for researching the properties of those cells in biology.

Furthermore, the basis formed by the vector cells collectively forms a grid cell's response. For example, plotting the four signals together as one cell in a tracking trail in figure 11 shows how this cell is triggered in a square-grid fashion. A hexagonal grid can be implemented simply by replacing the current orthogonal basis with 6 basis vectors separated by $60°$. Our model then provides an explanation of how grid cells tessellate. Interestingly, this explanation suggests that the grid cell's main function is to generate the phase reset signals for the theta cells.

We have hinted several times at the potential of the place cell network we proposed, especially the potential of the yet-to-exist direct connections between the place cells. Since the place cell network encodes the space in a coherent map, conventional path finding algorithms can be applied for a complete SLAM system. Any smart algorithm targeting at finding the optimized path requires some form of knowledge between the current location and the destination. We believe that the connection between place cells can broadcast distance information through the diffusion of the interconnections between place cells, given that the place cells can perform memory tasks. Then, a computed path can be expressed as a sequence of path cells by which to perform the navigation task. Further, a more dynamic navigation algorithm can be combined with the flexible spatial encoding capability of the place cell network through connection with vector cells with different magnitudes. These topics will be our group's future research toward a complete neuromorphic SLAM system.

In summary: we proposed and validated a neuromorphic system inspired by the hippocampus for spatial localization. The system can track the location of a moving agent under the influence of variations among the system with a simplified digital interference and still reproduce the behavior of place cells. Overall, the proposed system provides a concise and implementable platform for a more efficient neuromorphic SLAM system.